\documentclass[aps,twocolumn,pre,amsmath,amssymb]{revtex4-1}
\usepackage{graphicx}

\begin{document}
\title{Identifying rare chaotic and regular trajectories in dynamical systems with Lyapunov weighted path sampling}
\author{Philipp Geiger}
\author{Christoph Dellago}
\email{Christoph.Dellago@univie.ac.at}
\address{Faculty of Physics, University of Vienna, Boltzmanngasse 5, 1090 Vienna}

\begin{abstract}
Depending on initial conditions, individual finite time trajectories of dynamical systems can have very different chaotic properties. Here we present a numerical method to identify trajectories with atypical chaoticity, pathways that are either more regular or more chaotic than average. The method is based on the definition of an ensemble of trajectories weighted according to their chaoticity, the Lyapunov weighted path ensemble. This ensemble of trajectories is sampled using algorithms borrowed from transition path sampling, a method originally developed to study rare transitions between long-lived states. We demonstrate our approach by applying it to several systems with numbers of degrees of freedom ranging from one to several hundred and in all cases the algorithm found rare pathways with atypical chaoticity. For a double-well dimer embedded in a solvent, which can be viewed as simple model for an isomerizing molecule, rare reactive pathways  were found for parameters strongly favoring chaotic dynamics.  
\end{abstract}

\maketitle


\section{Introduction}
\label{sec:intro}

The phase space of dynamical systems often exhibit regions with qualitatively very different dynamics. In the Henon-Heiles model or similar low-dimensional Hamiltonian systems, for instance,  islands of stability are embedded in a chaotic sea \cite{Ott1993}. Other examples for this kind of behavior include the Fermi-Pasta-Ulam chain, in which special initial conditions lead to physically very different soliton and "chaotic breather" solutions, and gravitational many-body systems in celestial mechanics, which, for appropriate initial conditions, produce orbits that are stable for very long times \cite{Murray1999}. Although such particularly regular (or irregular) trajectories may be very rare, they may be responsible for important physical phenomena as is the case for chemical reactions, where trajectories passing through unstable saddle points regions carry the system from one chemical species to another \cite{Komatsuzaki2002,Wiggins2009}.

Identifying and describing such trajectories is therefore of great interest. Recently, Tailleur and Kurchan \cite{Kurchan2007} presented a powerful new method , the \textit{Lyapunov weighted dynamics (LWD)}, which is not limited to low dimensions or restricted to a small family of problems. In this evolutionary approach, a swarm of walkers progress according to the rules of the underlying dynamics. The walkers proliferate or die depending on the degree of chaos encountered by the system along a particular trajectory and, after many generations, only walkers on trajectories with the desired stability properties survive. Tailleur and Kurchan have demonstrated that their method is capable of finding even very small stability regions in systems of many degrees of freedom. 

Inspired by the work of Tailleur and Kurchan, we introduce here a general and efficient algorithm for finding trajectories with atypical stability properties, which is equally applicably to stochastic and deterministic dynamics. The central notion of our approach lies in the definition of a Lyapunov weighted path ensemble, in which the statistical weight of trajectories explicitly depends on a measure for the chaoticity of the underlying dynamics. The degree with which particularly chaotic trajectories are favored or disfavored depends on the value of a parameter that can be viewed as conjugate to the measure of chaos, for instance the Lyapunov exponent. This ensemble of trajectories is then sampled using techniques borrowed from transition path sampling (TPS), a method originally developed to study rare transitions between long-lived stable states in complex molecular systems \cite{Dellago1998,Bolhuis2000,Dellago2002,Dellago2008}. Note that a related combination of transition path sampling with a Lyapunov weighted action has been suggested before \cite{Kurchan2003,Kurchan2004}. By construction, the transition path sampling procedure generates a Markov chain of trajectories distributed according to the chosen bias function. Following the terminology of Tailleur and Kurchan \cite{Kurchan2007}, we call this approach {\em Lyapunov weighted path sampling}. Similar techniques were recently used by Chandler and collaborators to sample an ensemble of trajectories weighted by an order parameter describing the mobility of particles in a system undergoing the glass transition \cite{Chandler2009}.

The remainder of this article is organized as follows. In Sec. \ref{sec:methods} we briefly introduce the concept of Lyapunov instability and discuss ways to probe chaotic dynamics along individual trajectories. In Sec. \ref{sec:ensemble} we define the Lyapunov weighted path ensemble  and in Sec. \ref{sec:sampling} we explain how it can be sampled with transition path sampling algorithms. In Sec. \ref{sec:systems} this approach is then applied to detect particularly stable and unstable trajectories in various systems. Some conclusions are provided in Sec. \ref{sec:conclusions}.


\section{Lyapunov instablity}
\label{sec:methods}

Chaotic dynamical systems are characterized by a strong sensitivity to small changes in the initial conditions. To quantify this concept, consider a dynamical system, described by a set of $N$ coupled first order ordinary differential equations,
\begin{equation}
\label{Equ_system_evolution}
	\dot x = F(x),
\end{equation}
where $x$ denotes a point in the $N$-dimensional phase space. The time evolution of a infinitesimally small deviation $\delta x$ separating two close-by trajectories is then governed by the linearized equations of motion
\begin{equation}\label{Equ_deviation_vector_evolution}
 \delta  \dot  x(t) = D(x) \delta x (t),
\end{equation}
where $D(x) = \partial F/\partial x$ is the Jacobi matrix of the system evaluated at $x$. In a chaotic system, two points in phase space, initially separated by $\delta x (0)$ at time $t=0$, will lead to trajectories that, on the average, separate exponentially in time, $| \delta x(t) | \approx | \delta x(0) | \exp(\lambda t)$. Here, the vertical lines denote the Euclidean norm of a vector. The coefficient $\lambda$, the long-time averaged growth rate of infinitesimally small displacements defined as
\begin{equation}
\lambda = \lim_{t \rightarrow \infty} \frac{1}{t} \ln \frac{| \delta x(t) |}{| \delta x(0) |},
\label{equ:Lyapunov}
\end{equation}
is called the Lyapunov exponent of the system. A positive Lyapunov exponent corresponds to exponential growth of an initially small perturbation and implies information loss and strong sensitivity to initial conditions, the defining feature of chaotic dynamics. (It is possible to define whole spectra of Lyapunov exponents characterizing the growth rates of small perturbation in different directions of phase space \cite{Benettin1980,Posch1994}. In this article, however, we will consider only the largest Lyapunov exponent defined in Equ. (\ref{equ:Lyapunov}).)

Since Lyapunov exponents are defined as long-time averages, in an ergodic system every initial condition will yield the same $\lambda$. For finite periods of time however, trajectories can display very different chaotic properties. An example are trajectories in ``sticky regions'' in the phase space of the standard map, in which trajectories can spend a long time before escaping away into more chaotic regions \cite{Reinhardt1982,Voglis1998}. Other parts of phase space may be filled with regular periodic orbits that have a vanishing Lyapunov exponents and are dynamically disconnected with the chaotic surroundings.  To describe such behavior of individual trajectories of finite length it is convenient to consider so-called finite time Lyapunov exponents $\lambda_{f}({x_0}, t)$ that depend on the initial condition $x_0$ and on the temporal trajectory length $t$, 
\begin{equation}
\label{Equ_finite_time_lyapunov_exponent}
  \lambda_{f}(x_0, t) = \frac{1}{t} \ln{\frac{| \delta x(t)|}{| \delta x(0)|}}.
\end{equation} 
Such finite time Lyapunov exponents can be used to quantify the chaoticity of finite length trajectories.  

One difficulty occurring in the definition of the finite time Lyapunov exponent of Equ. (\ref{Equ_finite_time_lyapunov_exponent}), however, is that $\lambda_f$ also depends on the initial orientation of the displacement vector $\delta x(0)$. This vector reorients into the direction of fastest growth, but, depending on the degree of chaos prevalent in the respective phase space region, this reorientation may take a time long with respect to the trajectory length $t$. In fact, the time $\tau_r$ it takes to turn the displacement vector into the direction of the fastest growth is inversely proportional to the difference of the two largest Lyapunov exponents $\tau_r \propto 1/(\lambda_1 - \lambda_2)$ \cite{DellagoHooverPosch2002}. Here, $\lambda_1$ and $\lambda_2$ are the largest and the second largest Lyapunov exponent, respectively. This ambiguity in the definition of the finite time Lyapunov exponent can be avoided by integrating the equations of motion backwards for a time $t_-$ longer than $\tau_r$ starting from the inititial condition $x_0$. If the equations of motion for the displacement vector $\delta x$ are then integrated forward starting from $x_{-t_-}$ with an arbitrary orientation of the displacement vector and following the reference trajectory, the displacement vector has oriented into the direction of fastest growth when $x_0$ is reached. Then, the displacement vector has a well defined orientation at $t=0$ and the definition of the finite time Lyapunov exponent for the trajectory from $x_0$ to $x_t$ is unique. Since the reorientation time $\tau_r$ may be large, this procedure can require the computation of long additional trajectory segments causing large computational costs.

An alternative way to quantify the chaoticity of individual finite length trajectories consists in determining the \textit{Relative Lyapunov Indicator (RLI)} \cite{Sandor2000,Sandor2004}, originally introduced to detect chaotic dynamics in planetary systems. This measure has been proven particularly useful to distinguish regular from chaotic trajectories in systems that are only weakly chaotic. The main idea of the RLI is to exploit the fact that in the chaotic regions of phase space, finite time Lyapunov exponents vary discontinuously as a function of the initial condition, i.e.,~adjacent points can have very different local expansion rates \cite{Voglis1997,Dellago2000}. The RLI is defined as the magnitude of the difference $\Delta \lambda (x_0, t)$ between the finite time Lyapunov exponents of two trajectories separated by a small but finite amount $\Delta x_0$,
\begin{equation}
	\Delta \lambda(x_0,t) = \left| \lambda(x_0 + \Delta x_0, t) - \lambda(x_0, t) \right|.
\end{equation}
It has been shown that the RLI is insensitive both to the separation $\Delta x_0$ as well as to the initial infinitesimal phase space displacement $\delta x(0)$ used to calculate the finite time Lyapunov exponents $ \lambda(x_0, t)$. 
To reduce fluctuations, one can average the RLI over time,
\begin{equation}
	R(x_0, t) = \overline{ \Delta \lambda (x_0, \Delta t)} = \frac{1}{t/\Delta t} \sum_{i=1}^{t/\Delta t} \Delta \lambda(x_0, i\Delta t),
\end{equation}
where $\Delta t$ is the time step used for the numerical integration of the equations of motion. In the following we will use this smoothed version of the RLI to characterize the chaoticity of individual trajectories in various dynamical systems. 


\section{Lyapunov weighted trajectory ensemble}
\label{sec:ensemble}

As outlined in the Introduction, the goal of the work presented in this paper is to identify trajectories that have particular stability properties. For instance, one might be interested in locating regions of phase space that are populated by regular trajectories or in finding those pathways that are most chaotic. To do that, we start by defining an ensemble of trajectories including an additional weight that favors trajectories with the desired chaoticity properties. We assume that the dynamics generates a stationary distribution $\rho(x)$. Since we consider only deterministic systems here, we represent trajectories of length $t$ by their initial phase space point $x_0$ and define the {\em Lyapunov weighted path ensemble} (LWPE) as
\begin{equation}
\rho_L(x_0, t)=\frac{1}{Q}  \rho(x_0) e^{\alpha t R(x_0, t)},
\end{equation}
where the factor $Q=\int dx_0 \rho(x_0) e^{\alpha t R(x_0, t)}$ normalizes the distribution. Here, the parameter $\alpha$, which can be viewed as conjugate to the chaoticity indicator $R(x_0, t)$, controls how strongly the weight of initial condition $x_0$ is changed according to the chaoticity of the trajectory evolving out of $x_0$. An analogous ensemble of trajectories can be easily defined using finite time Lyapunov exponents if they can be calculated accurately (simply replacing $R(x_0, t)$ with $\lambda_f(x_0, t)$ in the above equation). Large positive values of $\alpha$ favor strongly chaotic trajectories with a large chaoticity indicator. For negative $\alpha$s, on the other hand, weakly chaotic or regular trajectories with a small chaoticity indicator are given a larger weight in the ensemble. Note that a similar ensemble of trajectories can easily be constructed also for stochastic dynamics and appropriately defined Lyapunov exponents.  


\section{Sampling trajectories with TPS}
\label{sec:sampling}

We sample the Lyapunov weighted path ensemble with a technique borrowed from transition path sampling \cite{Dellago1998,Dellago2002,Dellago2008}, a method originally developed to simulate rare but important transitions between long-lived stable states as they occur, for instance, in protein folding, chemical reactions and first order phase transitions. In a transition path sampling simulation a biased random walk is carried out in the space of trajectories in a way such that trajectories are sampled according to their weight in the desired ensemble. This is accomplished using a Monte Carlo procedure in which a trial trajectory is generated from the current trajectory and then accepted according to the Metropolis rule. Iterating this basic step, a set of trajectories with the correct probability is generated. A particularly efficient way to generate trial pathways is the so called shooting algorithm \cite{DellagoBC1998,Faraday1998}, also used in the present work. In this approach, a new trajectory is generated from an old one by first randomly selecting a point on the old trajectory, and then integrating the equations of motion starting with perturbed momenta.The magnitude of the perturbation of the momenta controls how different the new trajectory is from the old one and, therefore, also controls the average acceptance probability of the Monte Carlo procedure.  

In more detail, the path sampling procedure of the Lyapunov weighted path ensemble is carried out in the following way. The first trajectory is created by integrating the equations of motion starting from an arbitrary initial condition $x_0$. From this trajectory one then selects a point at random. At this so-called shooting point, the momenta are slightly changed by addition of a small perturbation drawn from a Gaussian distribution. The new trajectory is then obtained by integrating the equations of motion forward to time $t$ and backward to time $0$. The new trajectory with initial condition $x_0^{\rm (n)}$ and chaoticity indicator $R(x_0^{\rm (n)},t)$ is accepted with probability   
\begin{equation}
\label{equ_acc_prob}
	p_{\rm acc} = \min\left\lbrace 1, \frac{\rho(x_0^{\rm (n)})}{\rho(x_0^{\rm (o)})}e^{\alpha t \left[  R(x_0^{\rm (n)},t) - R(x_0^{\rm (o)}, t) \right] }\right\rbrace .
\end{equation}
where $x_0^{\rm (o)}$ and $R(x_0^{\rm (o)}, t) $ denote the initial point and the chaoticity indicator of the old trajectory, respectively. If the new trajectory is accepted, the procedure will be repeated with this trajectory. Otherwise the old trajectory is kept as the current one. The acceptance probability of Equ. (\ref{equ_acc_prob}) is derived from the detailed balance condition and guarantees that trajectories are harvested according to the Lyapunov weighted path ensemble. Note that to obtain Equ. (\ref{equ_acc_prob} ) we have assumed a momentum perturbation that leads to a symmetric generation probability. If this is not the case, an appropriate factor must be taken into account in the acceptance probability. The shifting algorithm of transition path sampling \cite{DellagoBC1998,Faraday1998} can be adapted in a similar way to sample the Lyapunov weighted path ensemble. In the next section we discuss the application of this method to various chaotic dynamical systems.


\section{Results}
\label{sec:systems}

In this Section we use the method outlined above to identify particularly stable and unstable trajectories in various dynamical systems with dimensionality ranging from two to several hundred. For better comparison, we mostly follow Ref. \cite{Kurchan2007} in our choice of examples. 

\subsection{Standard map}
\label{sec:systems_standard_map}

The standard map is a representation of the dynamics of a free rotor kicked at regular intervals with an impulsive force in a given direction \cite{Chirikov1979,Ott1993}. For a kicking period of 1 and a kick strength of $k$, the standard map is given by 
\begin{eqnarray}
  \omega_{n+1} &=& \omega_n - \frac{k}{2\pi} \sin(2\pi\varphi_n),  \nonumber \\
   \varphi_{n+1} &=& \varphi_n + \omega_{n+1}, 
  \label{label_standardmap}
\end{eqnarray}
where $\omega_n$ and $\varphi_n$ are the angle and the angular momentum of the rotor immediately after the $n$-th kick, respectively. Due to the periodicity of the motion, both variable $\omega$ and $\varphi$ are considered on a torus (by taking $\omega$ and $\varphi$ modulo 1). The standard map is area preserving and, depending on the value of $k$, displays different degrees of chaos. For vanishing $k$, the dynamics reduces to the motion of a free rotor and the system is integrable. Accordingly, the largest Lyapunov exponent vanishes. As the parameter $k$ is turned on, some chaos develops in particular regions of the two-dimensional phase space, while other regions originating from KAM-tori remain regular resulting in a phase space structure consisting of islands of stability embedded in a chaotic sea \cite{Chirikov1979,Ott1993}. As $k$ is increased, the stable regions shrink and for large values of $k$ only a very small fraction of initial conditions lead to regular trajectories. In the following we will sample the Lyapunov weighted path ensemble to find these rare regular trajectories. 

To sample trajectories of the standard map we use the standard shooting algorithm, in which a new trajectory is obtained from the current one by first selecting a phase space point on the current trajectory. Then, the selected point is slightly perturbed by addition of a random displacement drawn from a Gaussian distribution with width $\sigma_{\rm G}$ to both $\varphi$ and $\omega$. Starting form this perturbed initial condition, the new trajectory is obtained by carrying out an appropriate number of iterations of the standard map in forward direction and of the inverse map in backward direction. Finally, the new trajectory is accepted with the acceptance probability of Equ. (\ref{equ_acc_prob}). 

\begin{center}
\begin{figure}[t]
\includegraphics[width=7.0cm]{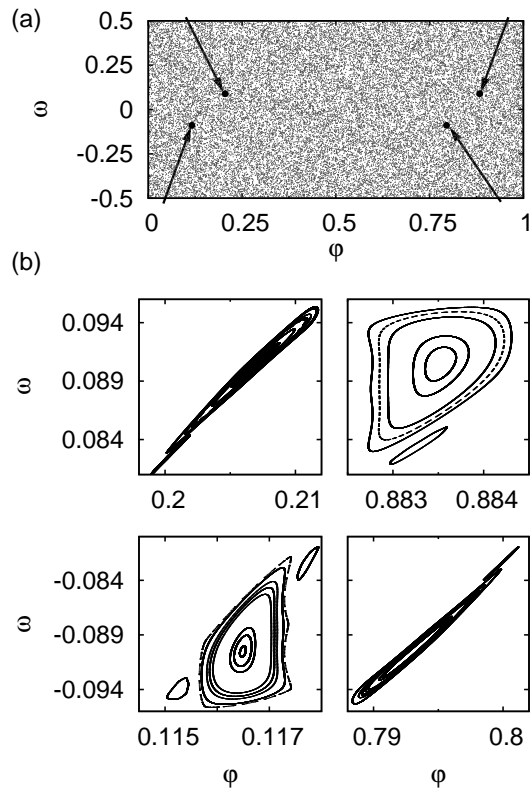}
\caption{(a) Phase space plot of the standard map for $k=7.7$. The arrows indicate the locations of the stability islands, which are not visible at the scale of the figure. (b) Enlarged phase space portraits of the regular trajectories found by the Lyapunov weighted path sampling algorithm.}
\label{fig:smap}
\end{figure}
\end{center}

Using the shooting algorithm we have generated $10^5$ trajectories of length $n=10^4$ with $\sigma_{\rm G}= 0.05$ for a kicking strength of $k=7.7$.  In this case, the chaotic sea covers almost the entire phase space. The magnitude of initial point deviation for the RLI calculation was $|\Delta x_0| = 10^{-12}$ and the probability density $\rho(x_0)$ of initial conditions was assumed to be uniform. To find the rare regular trajectories a negative value of the control parameter $\alpha=-4$ was used. The right choice of this parameters is crucial as it controls how phase space is sampled in a way that is analogous to the effect of the inverse temperature in a regular Monte Carlo simulation. A very large value of $\alpha$ favors regular trajectories very strongly such that the simulation gets trapped very easily in local minima of the RLI and the most regular trajectories are not found. A small value of $\alpha$, on the other hand, does not generate a sufficiently strong bias towards the atypical regular trajectories such they may not be found in this case either. Using these parameters, several rare regular trajectories were found in the simulation as displayed in Fig.~\ref{fig:smap}. These regular trajectories correspond to the four stable islands found also in Ref.~\cite{Kurchan2007}. For the stable orbits the RLI values ranged from $10^{-14}$ to $10^{-12}$ compared to values of $10^{-4}$ for typical trajectories in the chaotic sea. As can be inferred from Fig.~\ref{fig:smap}, the regular islands cover only a very small part of phase space. The fact that regular trajectories are nevertheless found in the simulation indicates that the landscape of the chaoticity indicator must have a global funnel-like shape that attracts the simulation towards the regions of highest regularity.

\subsection{Spring pendulum}
\label{sec:pendulum}

We next consider the spring pendulum evolving according to Hamilton's equation of motion. This two-dimensional system consists of a point of mass $m$ exposed to a constant force of magnitude $g$ in negative $y$-direction attached to a fixed pivot by a harmonic spring such that both angle and length of the pendulum can change in time. The Hamiltonian of the spring pendulum is given by
\begin{equation}\label{equ:pendulum_hamiltonian}
 \mathcal H = \frac{p^2}{2m} + \frac{k}{2} (r - R)^2 + g \, y, 
\end{equation}
where $r=\sqrt{x^2+y^2}$ is the distance of the mass point to the pivot, $p^2=p_x^2+p_y^2$ is the squared magnitude of the momentum, $k$ is the spring constant and $R$ is the equilibrium length of the spring \cite{Posch1990,Dellago2000}. In all following calculations, the mass $m$, the equilibrium length $R$ and the force constant $k$ are set to unity and the force strength to $g=2$. 

%
\begin{figure}[t]
 \centering
 \includegraphics[width=7.0cm]{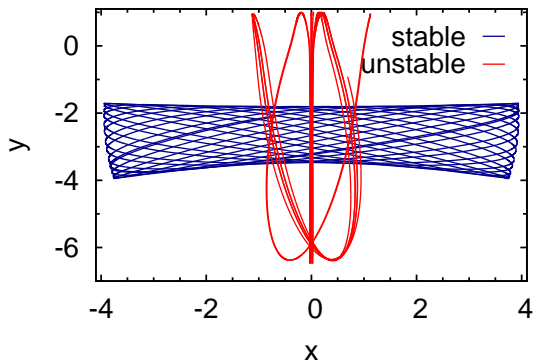}
 \caption{Very stable and unstable trajectories of the spring pendulum traced out in configuration space. In both cases the energy is $E=2$.}\label{fig:pendulum}
\end{figure}
%

For this model, we have carried out a Lyapunov weighted path ensemble simulation for two different values of the parameter $\alpha$. In both cases, the total energy was $E=2$ and the equations of motion are integrated with a time step of $\Delta t = 10^{-3}$ using the symplectic time-reversible Forest-Ruth algorithm of fourth order \cite{Omelyan2002} yielding good energy conservation. The shooting algorithm was carried out in the usual way by adding perturbations drawn from a Gaussian distribution to the momenta at the shooting point. Results of these simulations are displayed in Fig.~\ref{fig:pendulum}. The very stable trajectory of length $\tau=100$ was obtained after $4000$ iterations of the path sampling algorithm with $\alpha=-5\times 10^{11}$ favoring regular trajectories. In this calculation the magnitude of the shooting displacement was $\sigma_{\rm G}= 0.5$. This trajectory has an RLI of $\sim 2 \times 10^{-14}$, a value which is about four orders of magnitude smaller than that of a typical trajectory.

The unstable trajectory of Fig.  ~\ref{fig:pendulum} has been found after $3000$ iterations carried out with $\alpha=10^6$ and a shooting displacement of magnitude $\sigma_{\rm G} = 5\times 10^{-4}$. This trajectory has an RLI of $\sim 6 \times 10^{-4}$, which is about 6 orders of magnitude larger than that of typical trajectories. The qualitative difference between the two trajectories of Fig.  ~\ref{fig:pendulum} is remarkable. While the stable trajectory has much amplitude in the angular degree of freedom, the unstable one displays pronounced stretching movements with smaller angular oscillations. 

\subsection{Fermi-Pasta-Ulam chain}
\label{sec:systems_fpu}

As an example of a system with higher dimensionality, we search for particularly stable trajectories of the Fermi-Pasta-Ulam (FPU) chain, originally concocted to study the thermalization of oscillatory modes in solids \cite{Fermi1955,Ruffo2005}. This model consists  of a one-dimensional chain of $N$ point particles with mass $m$ located at positions $x_i$. The particles are coupled by harmonic springs to which a weak anharmonic part in form of a quartic potential is added leading to the Hamiltonian
\begin{equation}
\label{equ:fpu_hamiltonian}
 \mathcal H = \sum_{i} \frac{p_i^2}{2m}  + \sum_i \frac{k}{2} (x_{i+1}-x_i)^2 + \frac{\beta}{4}  (x_{i+1}-x_i)^4.
\end{equation}
Here, $k$ is the spring constant of the harmonic spring and $\beta$ is a parameter controlling the strength of the quartic potential. Below, we use units in which $m=1$, $k=1$ and consider the case $\beta=0.1$ with fixed boundary conditions. 

\begin{figure}[t!]
 \centering
 \includegraphics[width=8.0cm]{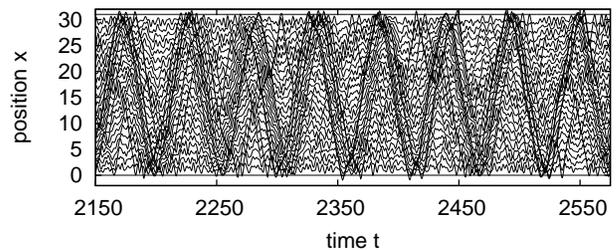}
 \caption{Particle positions of a FPU-chain with $N=32$ as a function of time along part of a trajectory sampled by the Lyapunov weighted path sampling algorithm. Particle positions have been displaced vertically for better visibility. The trajectory was obtained after a couple of thousand iterations with $\alpha=-5\times10^8$ favoring particularly stable trajectories. }\label{fig:fpu}
\end{figure}

Depending on initial conditions, the FPU-chain displays different types of motion. Starting from a state in which all the energy is concentrated in a high frequency mode, the system evolves into a so-called breather moving chaotically with the energy strongly localized in space before equilibration eventually sets in on very long time scales \cite{Torcini1998}. Other initial conditions lead to only weakly chaotic solitonic modes, in which a kink moves through the system with constant speed and a preserved shape of its sharp front \cite{Kruskal1965,Ruffo2005}. Here, we use our algorithm to find these weakly chaotic solitons by using a parameter $\alpha=-5\times10^8$ highly favoring regular trajectories. Note that this strongly negative value of $\alpha$ was necessary since chaos in the FPU-chain in this regime is very weak and it is difficult to distinguish between more and less chaotic trajectories. Particle traces for a system of $N=32$ particles at a total energy of $E = 32$ are shown in Fig.~\ref{fig:fpu}. The equations of motion were integrated with a time step of $\Delta t = 0.05$ for trajectories of length $\tau=5000$. The trajectory shown in Fig.~\ref{fig:fpu} was obtained after a couple of thousand iterations of the path sampling scheme with shooting displacements in momentum space of magnitude $\sigma_{\rm G}=0.5$. Kinks wandering through the chain at constant velocity and bouncing back and forth between the chain ends are clearly visible indicating a solitonic mode of motion.

\subsection{Double well system}
\label{sec:systems_separatrix}

To test whether Lyapunov biased path sampling can be used to find reactive trajectories in systems with multiple stable states separated by barriers, we have studied a simple double well system in one dimension. The dynamics of the system is governed by the  Hamiltonian
\begin{equation}
\label{equ:hamiltonian_separatrix}
  H(x, p) = \frac{p^2}{2m} + k(x^2-1)^2,
\end{equation}
corresponding to a particle of mass $m$ moving on a potential energy surface with minima at $x = \pm 1, y=0 $ and a maximum at the origin. The potential energy barrier separating the two stable states has a height of $k$. We imagine that the system is in contact with a heat bath with temperature $T$, such that the distribution of initial conditions of the trajectories is canonical, $\rho(x, p)\propto \exp\{-\beta H(x, p)\}$. Here, $\beta = 1 /k_{\rm B}T$ is the reciprocal temperature and $k_{\rm B}$ is the Boltzmann constant. Thus, all initial conditions of the system are in principle accessible in this ensemble, albeit with different statistical weight. This system is integrable and hence its Lyapunov exponent vanishes. Nevertheless, finite length trajectories diverge strongly near the barrier top.

Although a canonical distribution of initial conditions implies a coupling of the dynamics to the degrees of freedom of the heat bath, we assume that this coupling is so weak that it does not affect the dynamics of the system on the time scale of the length $\tau$ of the trajectories. Hence, each individual trajectory evolves at constant total energy. Under these conditions, the acceptance probability of the path sampling procedure is given by
\begin{equation}
\label{Eqn_acc_prob_bisable_sys}
	p_{\rm acc} = \min\left\lbrace 1, e^{\alpha \tau \left[  R({\rm n},\tau) - R({\rm o}, \tau) \right]} e^{-\beta \left[H({\rm n})-H({\rm o})\right]} \right\rbrace,
\end{equation}
where $H({\rm n})$ and $H({\rm o})$ are the energies of the new and the old trajectories, respectively, and $R({\rm n},\tau)$  and $R({\rm o}, \tau)$ are the respective RLIs. The above acceptance probability takes into account possible energy changes due to the momentum perturbation applied at the shooting point.

\begin{figure}[t!]
 \centering
 \includegraphics[width=6cm]{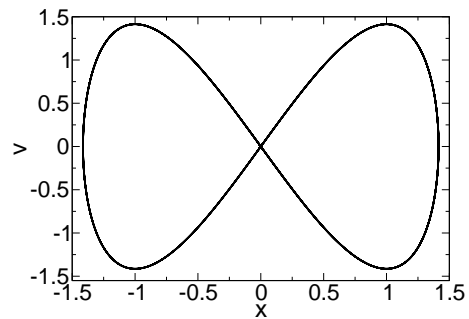}
 \caption{Phase space plot of a particularly chaotic trajectory crossing the saddle point in the double well system. Note that at the origin there is a gap between the upper and the lower branch of the periodic trajectory, but the gap is too small to be visible at the scale of the figure.}
 \label{fig:sep1}
\end{figure}

Since we are interested in reactive trajectories that cross the energetic barrier between the stable states and the phase space structures near saddle points which lie at origin of chaos \cite{Ott1993}, we sample the Lyapunov weighted path ensemble procedure with a large value of $\alpha=5 \times 10^8$ strongly favoring chaotic trajectories. The equations of motion are integrated with a time step of $\Delta t = 10^{-2}$ and trajectories have a total temporal length of $\tau = 10^2$. Units where chosen such that $m=1$ and $k=1$ and the temperature was set to $\beta=1$. The magnitude of the shooting displacement to the momenta was $\sigma_{\rm G} = 0.05$. Starting from a trajectory of energy $E=0.05$ oscillating about the bottom of one well, after a few hundred path sampling steps reactive trajectories connecting the two wells were obtained. The phase space plot of a reactive trajectory with total energy $E=1.0000019$ (just slightly above the barrier height of $E=1.0$) crossing the barrier in close proximity of the saddle point is shown in Fig.~\ref{fig:sep1}.  

\subsection{Double well dimer in a solvent}
\label{sec:system_dimer}

As exemplified by the results of the previous section, the dynamical character of pathways crossing barriers in the vicinity of saddle points in the potential energy surface strongly differ from that of trajectories fluctuating about minima  \cite{Komatsuzaki2002}. Such pathways connecting stable states are, for instance, relevant in the context of activated chemical reactions and there is intense interest in computational methods for finding such rare barrier crossing pathways along which chemical reactions occurs \cite{Dellago2008}. In this section we will study if the Lyapunov weighted path sampling method can be used to identify reactive trajectories based on their chaoticity. In particular, we will address the question wether such an approach can be successful for reactions occurring in solution, where the chaoticity of the reactive subsystem may be overshadowed by that of the solvent. 

 \begin{figure}[t]
\includegraphics[width=8cm]{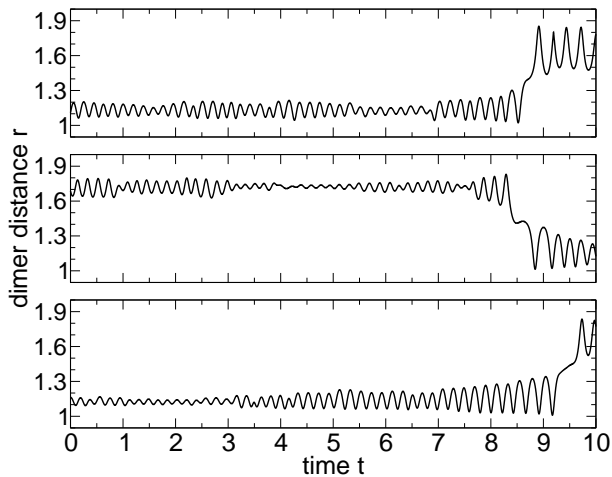}
 \centering
 \caption{Interatomic distance $r$ of the double well dimer as a function of time $t$ along reactive trajectories found by the Lyapunov weighted path sampling procedure for particle numbers $N=22$ (top), $N=32$ (center), and $N=108$ (bottom). In all three trajectories a transition from the contact to the extended state occurs as evidenced by the sudden increase of $r$ towards the end of the trajectory. }
\label{fig:dimer}
\end{figure}

We study this issue using a simple two-state model dimer embedded in a soft-sphere solvent \cite{Straub1988, Dellago1999}. This three-dimensional model consists of $N$ particles of mass $m$ evolving according to Hamilton's equations of motion in a cubic box of volume $V$ and periodic boundary conditions. All particle interact pairwise via the purely repulsive Weeks-Chandler-Andersen (WCA) potential \cite{Weeks1971},
\begin{equation}
V_{\rm WCA}(r) = \left\{\begin{array}{ll}
4\varepsilon \left[ \left(\frac{\sigma}{r}\right)^{12} - \left(\frac{\sigma}{r}\right)^{6} \right] + \varepsilon & \mbox{ for } r \leq r_c, \\
0 & \mbox{ for } r > r_c.
\end{array}\right.
\end{equation}  
Here, $r$ is the interparticle distance, $\sigma$ is the interaction radius, $\varepsilon$ is the strength of the potential, and $r_c = 2^{1/6}\sigma$ is the cutoff radius. In addition, the two particles forming the dimer are bonded through the double well potential 
\begin{equation}
  V_{\rm dw}(r) = h \left[ 1 - \frac{(r-r_c-w)^2}{w^2} \right]^2,
\end{equation}
where $h$ and $w$ are constants determining the height and the width of the barrier, respectively. The distance between the two minima of the double well potential is $2w$. For low temperatures, i.e., for $h \gg k_{\rm B} T$, the dimer mainly exists in two states. In one state, the contact state, the interatomic distance of the dimer fluctuates about $r_c$. In the other state, the extended state, the dimer has a bond length of about $r_c + 2w$. Thermally activated transitions between these two states occur very rarely and are separated by long permanence times in the wells. 

In all our simulations we use reduced units in which $\sigma=1$, $\varepsilon=1$, and $m=1$. We study this system for particle numbers $N=22$, $N=32$ and $N=108$ at a density of $\rho=N/V=0.5$ and at a total energy per particle of $E/N=1.0$ corresponding to a temperature of about $T=0.55$. The dimer barrier height was set to $h=6.0$ and its width to $w=0.3$. Trajectories of total length $\tau=10$ were integrated with a time step of $\Delta t = 10^{-3}$. In the path sampling procedure, different parameters $\alpha$ and different magnitudes $\sigma_{\rm G}$ of the shooting displacements were used. While for the smallest system $\alpha=170$ and $\sigma_{\rm G}= 0.2$ were employed, we chose $\alpha=160$ and $\sigma_{\rm G} = 0.25$ for the system of intermediate size and $\alpha=600$ and $\sigma_{\rm G} = 0.05$ for the largest system. Starting from non-reactive trajectories, the Lyapunov weighted path sampling algorithm carried out with these parameters succeeded in finding reactive trajectory. Examples of the interatomic distance as a function of time are shown in Fig.~\ref{fig:dimer} along reactive trajectories for various system sizes. While for all system sizes studied here the Lyapunov weighted path sampling simulation converged towards reactive trajectories, the number of iterations before observing the first reactive path increased with system size. While for particle number $N=22$ the first reactive event occurred after about $400$ path sampling steps, the first reactive trajectory was found after $1000$ iterations for $N=32$ and after $2900$ iterations for $N=108$. These results indicate, that the pronounced chaoticity of barrier crossing trajectories can be used to identify reactive pathways even for systems with hundreds of degrees of freedom. Note, however, that not all reactive trajectories are strongly chaotic. During our simulations it happened repeatedly that reactive pathways were rejected, because their relative Lyapunov indicators were too small. It could be that along these trajectories the chaoticity associated with the saddle point crossing was compensated by a particularly stable dynamics of the solvent.


\section{Conclusions}
\label{sec:conclusions}

In this paper, we have presented a flexible numerical method to find particularly chaotic or regular trajectories in dynamical systems. The basic idea of the method, inspired by the work of Tailleur and Kurchan \cite{Kurchan2007} and called Lyapunov weighted path sampling, is to first define an ensemble of trajectories weighted by a measure of their chaoticity, for instance their finite time Lyapunov exponent. In this trajectory ensemble a parameter, which can be viewed as conjugate to the Lyapunov exponent, can be tuned to favor either very chaotic or very regular trajectories. The trajectory ensemble is then sampled with methods adopted from transition path sampling. Other chaoticity indicators besides finite time Lyapunov exponents can be easily integrated into the algorithm as well. Since the calculation of finite time Lyapunov exponents can be computationally demanding, we have, for instance, used relative Lyapunov indicators (RLI) to bias trajectories according to their level of chaos. These indicators are particularly sensitive and are capable of distinguishing weakly chaotic trajectories from regular ones. While in this paper we have used Lyapunov weighted path sampling to study only systems evolving deterministically, the method can be applied as easily to stochastic dynamics provided an appropriate chaoticity indicator is available.

The complexity of the examples studied here ranges from a simple one-dimensional double well system to the FPU-model and a bistable dimer in a solvent with hundreds of degrees of freedom. In all cases, the Lyapunov weighted path sampling algorithm successfully identified trajectories with atypical chaoticity properties. While for the FPU-model we used Lyapunov weighted path sampling to find weakly chaotic solitonic modes of motion, we concentrated on highly chaotic trajectories for the dimer in solution. The results obtained for this simple model of a chemical reaction indicate that it is possible to use Lyapunov weighted path sampling to find rare reactive trajectories that pass through saddle points in the potential energy surface as they connect long-lived stable states with each other. Further studies will be necessary to clarify to which degree identifying such trajectories is made difficult by the chaos arising from degrees of freedom not directly coupled to the reaction (for instance solvent degrees of freedom) and possibly eclipsing the dynamical instability of the reactive subsystem.  
It will also be interesting to investigate whether chaoticity indicators such as the maximum Lyapunov exponent, the Kolmogorov-Sinai entropy or the relative Lyapunov indicators used in the present study correlate with the measures of mobility used by Chandler and collaborators to link the glass transition with a first order phase transition in trajectory space \cite{Chandler2009}. In their work, these authors started from the equilibrium distribution of pathways and added to it a bias that favors trajectories with low dynamical activity.  Chandler an coworkers demonstrated numerically that this transition displays all the features of a first-order transition occurring in trajectory space. It would be interesting to study if an analogous bias based on chaoticity indicators also leads to an equivalent first order transition in path space.  In such research it may be fruitful to combine Lyapunov weighted path sampling with advanced equilibrium simulation methods such as umbrella sampling \cite{Umbrella1977}, metadynamics \cite{Meta2002}, or parallel replica sampling \cite{Replica1995} directly acting on chaoticity indicators. 

\section*{Acknowledgments}

This work was supported via a Junior Research Fellowship of PG at the Erwin Schr\"{o}dinger International Institute for Mathematical Physics (ESI) in Vienna and by the Austrian Science Fund (FWF) under Grant P20942-N16. The calculations were carried out, in part, on the Vienna Scientific Cluster (VSC).


\end{document}